\documentclass[12pt]{article}

\author{Andrei Yu. Khrennikov\\
Center for Mathematical
Modeling \\ in Physics and Cognitive Sciences,\\
University of V\"axj\"o, S-35195, Sweden\\
Email:Andrei.Khrennikov@msi.vxu.se}

\title{Brain as Quantum-like Machine for Transferring Time into
Mind}

\begin{document}

\maketitle

\begin{abstract}
We propose a model of processing of information in the brain which
has the following distinguishing features: a). It is quantum-like
(QL). The brain uses the quantum rule (given by von Neumann trace
formula) for calculation of averages for psychological functions.
b). Those functions are considered as self-observations of the
brain. c). The QL-representation has the temporal basis. The brain
is a machine transferring time into cognition. d). Any cognitive
process is based on (at least) two time scales:  precognitive time
scale (which is very fine) and cognitive time scale (which is
essentially coarser).

To couple our model to physiology, behavioral science, and
psychology, we consider a number of known fundamental time scales in
the brain. Although the elaboration of those scales was based on
advanced experimental research, there are still many controversial
approaches and results. The temporal structure of the brain
functioning is very complex.
\end{abstract}

\section{Introduction}

The idea that the description of brain functioning, cognition, and
consciousness could not be reduced to the theory of neural networks
and dynamical systems (cf. Ashby (1952), Hopfield (1982), Amit
(1989), Bechtel and Abrahamsen (1991), Strogatz (1994),  van Gelder
(1995), van Gelder and Port (1995), Eliasmith (1996)) and that
quantum theory may play an important role in such a description has
been discussed in a huge variety of forms, see e.g. Whitehead (1929,
1933, 1939), Orlov (1982),
 Healey (1984), Albert and Loewer (1988, 1992),
 Lockwood (1989, 1996), Penrose (1989, 1994),
 Donald (1990, 1995, 1996), Jibu and Yasue (1992, 1994),
 Bohm and Hiley (1993), Stapp (1993),
 Hameroff (1994, 1998), Loewer (1996), Hiley and Pylkk\"anen (1997),
 Deutsch (1997), Barrett (1999),
 Khrennikov (1999, 2000, 2002, 2003, 2004, 2006a), Hiley (2000), Vitiello
 (2001), Aerts, D. and  Aerts S. (2007), Conte et al. (2007) and literature thereby.

This idea that quantum mechanics might have some consequences for
cognitive science and psychology was discussed at many occasions
already by fathers of quantum theory. We can mention, for example,
attempts of Niels Bohr to apply the quantum principle of
complementarity to psychology (see A. Plotnitsky 2001, 2002, 2007
for discussions). We can also mention the correspondence between
Pauli and Young about analogy between quantum and mental processes.

During the last 30 years it was done  a lot for the realization of
the very ambitious program of quantum reductionism. There were
various attempts to reduce mental processes to quantum physical
processes in the brain. Here we point out to fundamental works
Hameroff (1994, 1998) and Penrose (1989, 1994, 2005).

However, the quantum formalism provides essentially more
possibilities for modeling of physical, biological, and social
processes. One should distinguish quantum mechanics as physical
theory and its formalism. In principle, there is nothing surprising
that a formalism which was originally developed for serving to one
special physical theory can be used in other domains of science. For
example, we are not surprised that differential calculus which was
developed to serve to classical Newtonian mechanics was later used
in field theory, quantum mechanics, biology, economics. Nobody
protests against applying the classical probability calculus (the
Kolmogorov measure-theoretic model) to modeling of financial
processes and so on. In the same way one might import into cognitive
science and psychology the mathematical formalism of quantum
mechanics, even without trying to perform a reduction of mental
processes to quantum physical processes.

To escape misunderstanding, we shall reserve notations classical and
quantum for physics. And in applications outside physics we shall
use notations classical-like (CL) and quantum-like (QL).

By using non-reductionist QL-models one can escape some fundamental
problems arising in the quantum reductionist approach, e.g., the
presence of the huge gap between the quantum (physical) and
neurophysiological scales. However, the problem of coupling with
physical reality could not be just forgotten. Suppose that the
quantum processes in the brain as a physical system are not
responsible for mental phenomena. The natural question arises: "What
is then the mechanism (physical, chemical, biological) inducing the
QL-rules of mental processing?" In the present paper we shall show
that the temporal structure of the brain functioning could be
responsible for the QL-structure of processing of mental
information.

Our starting point is a series of works Khrennikov (2005a, b,
2006b-d) on a new interpretation of quantum mechanics as a special
representation of classical statistical mechanics. In such an
approach the quantum formalism is merely a way of representation of
information about systems (physical as well as biological). Suppose
that we are not able to collect the complete set of information
about a system (e.g., because of some restrictions for measurement
procedures and technologies). In such a situation we may,
nevertheless, try to create a model of phenomena which is based on
ignorance of a part of information. By our interpretation the
quantum formalism provides the consistent rules for such a modeling.

In this paper we shall apply methods developed in Khrennikov (2005a,
b, 2006b-d) to cognitive science and psychology. We are especially
interested in the following fundamental question: How can such a
QL-projection of information be realized in biological systems?

We propose a model of processing of information in the brain which
has the following distinguishing features:

\medskip

a). It is quantum-like (QL). The brain uses the quantum rule (given
by von Neumann trace formula) for calculation of averages for
psychological functions.

b). Those functions are considered as self-observations of the
brain.

c). The QL-representation has the temporal basis. The brain is a
machine transferring time into cognition.

d). Any cognitive process is based on (at least) two time scales:
precognitive time scale (which is very fine) and cognitive time
scale (which is essentially coarser).

\medskip

To couple our model to physiology, behavioral science, and
psychology, we consider a number of known fundamental time scales in
the brain. Although the elaboration of those scales was based on
advanced experimental research, there are still many controversial
approaches and results. The temporal structure of the brain
functioning is very complex. As {\it the physiological and
psychological experimental basis} of our QL-model we chosen results
of investigations on one special quantal temporal model of mental
processes in the brain, namely, {\it Taxonomic Quantum Model} --TQM,
see Geissler et al (1978), Geissler and Puffe (1982), Geissler
(1983, 85, 87,92), Geissler and Kompass (1999, 2001), Geissler,
Schebera, and Kompass (1999). The TQM is closely related with
various experimental studies on the temporal structure of mental
processes, see also Klix and van der Meer (1978), Kristofferson
(1972, 80, 90), Bredenkamp (1993), Teghtsoonian (1971). We also
couple our QL-model with well known experimental studies, see, e.g.,
Brazier (1970), which demonstrated that there are well established
time scales corresponding to the alpha, beta, gamma, delta, and
theta waves; especially important for us are results of Aftanas and
Golosheykin (2005), Buzsaki (2005).

The presence of fine scale structure of firing patterns which was
found in Luczak et al (2007) in experiments which demonstrated
self-activation of neuronal patterns in the brain is extremely
supporting for our QL-model.\footnote{"Even in the absence of
sensory stimulation, cortex shows complex spontaneous activity
patterns, often consisting of alternating "DOWN" states of
generalized neural silence and "UP" states of massive, persistent
network activity. To investigate how this spontaneous activity
propagates through neuronal assemblies in vivo, we recorded
simultaneously from populations of 50-200 cells in neocortical layer
V of anesthetized and awake rats. Each neuron displayed a virtually
unique spike pattern during UP states, with diversity seen amongst
both putative pyramidal cells and interneurons, reflecting a complex
but stereotypically organized sequential spread of activation
through local cortical networks. The timescale of this spread was
~100ms, with spike timing precision decaying as UP states
progressed," see Luczak et al (2007).} Of course, not yet everything
is clear in neurophysiological experimental research, see Luczak et
al (2007): "The way spontaneous activity propagates through cortical
populations is currently unclear: while in vivo optical imaging
results suggest a random and unstructured process Kerr et al (2005),
in vitro models suggest a more complex picture involving local
sequential organization and/or traveling waves, Cossart et al
(2003), Mao (2001), Ikegaya (2004),  Sanchez-Vives and McCormick
(2000), Shu,  Hasenstaub, and McCormick(2003), MacLean (2005)."

In any event our QL-model for brain functioning {\it operates on
time scales which are used in neurophysiology, psychology and
behavioral science. This provides an interesting opportunity to
connect the mathematical formalism of quantum mechanics with
theoretical and experimental research in mentioned domains of
biology.} We hope that our approach could attract the attention of
neurophysiologists, psychologists and people working in behavioral
science to quantum modeling of the brain functioning. On the other
hand, our QL-model might stimulate theoretical and experimental
research on temporal structures of the brain functioning.

\section{Quantum-like processing of
incomplete information}

As was pointed out, in this paper we consider not the quantum
mechanics -- a special physical theory which is applicable for a
special class of physical systems (so called quantum systems),-- but
its formalism -- a special mathematical formalism for representation
of information. The quantum formalism is a special way of processing
of incomplete information. However, if information cut off were done
occasionally, one would have a chaotic information picture. The
quantum formalism provides a possibility to create a consistent
processing through the information projection. Such a formalism was
first time found in physics at the beginning of 20th century.

Of course, our interpretation of quantum mechanics -- as an
incomplete description of quantum systems -- contradicts to the
original views of Bohr, Heisenberg, Pauli, von Neumann, Dirac and
many others who postulated that quantum mechanics is a complete
theory: the wave function provides the complete representation of
statistical information about a system, e.g., electron. However, our
``incomplete information processing interpretation'' might be
sympathetic for Einstein, Schr\"odinger, De Broglie, Bohm, Margenau,
Popper, and nowadays Marshal, Ballentine, De Baere, De Muynck,
Santos, Khrennikov and many others; cf. also with Svozil, 2006.

Even if the first application of processing of incomplete
information on the basis of the quantum formalism was found in
physics, there are no fundamental reasons to restrict its
applications only to physics. We are interested in applications to
cognitive sciences.

One might guess that the ability for the {\it quantum-like} (QL)
processing of information was developed by biological organisms.
From the very beginning of evolution biological organisms operated
with huge information flows. They could create a representation of
external world which was based on an information-projection such
that cuts of information flows were done in a consistent way. In the
process of evolution there could be developed the ability to work
with information  by using the QL-representation.

\medskip

We start with physics and we consider two time scales Khrennikov
(2006d). One scale, we call it {\it prequantum,} is a fine time
scale, another, we call it {\it quantum,} is  a coarser time scale.
Oscillations at the prequantum time scale are averaged and used for
probabilistic reasoning at the quantum scale. The latter time scale
is considered as {\it an observational time scale.}

It is important  to mention that it was shown mathematically that
one can really derive quantum averages as approximations of
classical averages at the prequantum time scale, Khrennikov (2006d).

In the conventional quantum mechanics for physical systems the two
time scale representation has a ``semi-subjective character.'' On
the one hand, the quantum time scale -- the atom time scale in
Khrennikov (2006d):
$$
t_q\approx 10^{-21} \rm{sec},
$$ and the prequantum time scale -- the
Planck time scale in Khrennikov (2006d):
$$
s_{pq}\approx 10^{-44} \rm{sec},
$$
are scales of real physical processes.

On the other hand, the choice of the quantum (observational) scale
and, hence, the concrete application of the quantum representation
of information is a consequence of the presence of a special class
of observers -- human beings -- and the special level of development
of measurement technologies.

\medskip

We now suppose that a biological system might create the
QL-representation and QL-processing of information which are based
on operating at two time scales.\footnote{Thus discovery  of quantum
mechanics for physical systems was simply a rediscovery of the basic
representation of information in the human brain!?} There is an
analogue of the prequantum time scale. Information which is
processed at that time scale is considered as {\it non-cognitive.}
Thus this is a time scale of subconsciousness. We call this time
scale precognitive and denote the {\it precognitive time } by $s$
and its scale unit by $s_{\rm{pc}}.$ There is also an analogue of
the quantum time scale. One can say that it is the observational
time scale. However, the crucial difference from the conventional
quantum mechanics (for physical systems) is that there are no
external observers. The brain performs observations on itself. It is
better to speak about {\it self-observational time scale.} It is
assumed that this is the time scale of cognition. We call it the
{\it cognitive time} scale and denote the cognitive time by $t$ and
its scale unit by $t_{\rm{c}}.$ Of course, we have the inequality:
$$
s_{\rm{pc}} < t_{\rm{c}}.
$$
The crucial parameter that determines the measure of
quantumness (or better to say QL-ness) of cognition  is the parameter:
\begin{equation}
\label{TC} \kappa= \frac{s_{\rm{pc}}}{t_{\rm{c}}}.
\end{equation}
It provides a numerical measure of deviation of the QL (fuzzy, unsharp)
representation of information from the ``classical'' (complete, sharp) one.

Under the assumption that the precognitive time scale $s_{\rm{pc}}$
is fixed, we find that for small periods of fluctuations
$t_{\rm{c}}$ the parameter $\kappa$ is very large. Thus {\it higher
frequencies (at the cognitive time scale) induce larger deviations
from the (complete) CL-processing of information.}

Huge amounts of information which are processed at the precognitive
time scale are neglected, but not arbitrary (randomly). There is the
QL-consistency in the information processing. Consequently, for low
frequencies (oscillations with long periods) this coefficient is
small. Therefore the QL-processing does not imply large deviations
from the CL-computational regime.

The crucial problem is to find those biological time scales which
induce the QL-representation of information. There are many ways
to create such time scales. We split the problem into the two parts:

\medskip

1) to find the precognitive time scale;

2) to find the cognitive time scale.

\medskip

It seems that (as in physics) the first problem is more complicated.
First we consider the second one. We start  the discussion on the
choice of the cognitive time scale in by considering experimental
evidences, see, e.g., Khrennikov (2006a) for discussion and
references, that a moment in {\it psychological time} correlates
with $\approx 100$ ms of physical time for neural activity. In such
a model the basic assumption is that the physical time required for
the transmission of information over synapses is somehow neglected
in the psychological time. The time ($\approx 100$ ms) required for
the transmission of information from retina to the inferiotemporal
cortex (IT) through the primary visual cortex (V1) is mapped to a
moment of psychological time. It might be that by using
$t_{\rm{c}}=100 \rm{ms},$ we shall get the right cognitive time
scale.

However, the situation is not so simple even for the second problem.
There are experimental evidences that the temporal structure of
neural functioning is not homogeneous. The time required for
completion of color information in V4 ($\approx 60$ ms) is shorter
that the time for the completion of shape analysis in IT ($\approx
100$ ms). In particular it is predicted that there will be under
certain conditions a rivalry between color  and form perception.
This rivalry in time is one of manifestations of complex level
temporal structure of brain.

\medskip

Our fundamental assumption is that {\it there exist various pairs of
scales inducing various QL-representations of information.} In the
next section we shall discuss such a temporal QL-model of cognition
in more detail.

We shall come back to the ``difficult problem, namely, determination
of the precognitive time scale, in section 5. But at the moment we
forget about physiological and psychological time scales in the
brain and we present in more detail our QL-approach for processing
of information.

\section{Quantum-like approximation of temporal statistical averages in brain}

There are two time scales, a precognitive time scale $s_{\rm{pc}}$
and a cognitive time scale $t_{\rm{c}}.$ There is a cognitive
process $\pi$ (e.g., a cognitive task) which is performed at the
$t_{\rm{c}}$-scale. It integrates a number of processes which are
performed at the $s_{\rm{pc}}$-scale. Here "integrate" has the
meaning to produce averages with respect to oscillations at the
$s_{\rm{pc}}$-scale. Such averages are considered as cognitive
quantities at the level of the $\pi$-process.

In our model {\it "self-observation" is nothing else than the
calculation of an average.} However, this is only a part of the
story. If the brain were compute averages by the CL-algorithm -- as
statistical sums (with respect to huge ensembles of oscillations at
the precognitive time scale) -- then it would be simply an analogue
of the ordinary computer. This would be a kind of "statistical
physics thinking."

In our approach the QL-story of processing of information is in fact
the purely computation story.\footnote{No quantum mysteries at all.
The basic point is evolution for optimization of computational
abilities. The QL-computation is quicker, because it is an
approximative and because it does not need so much resources as the
CL-computation.} To calculate averages as statistical sums (over
huge neuronal ensembles), the brain should consume too much
computational and, hence, physical resources. My guess is that the
brain has the ability to perform calculations of averages by using
the rules of quantum mathematical formalism. Instead of a huge
statistical sum, the brain calculates its QL-approximation  given by
the von Neumann formula for the quantum average given by the
operator trace, see von Neumann (1955).

A classical mental quantity (psychological function) is given by a
function $f(\omega)$depending on the vector of parameters $\omega$
which are produced at the precognitive time scale. In the
QL-algorithm $f$ is approximated by its second derivative,
Khrennikov (2005a,b, 2006b-d). In this way the brain obtains a
symmetric operator $A,$ Hessian of the map $f.$ This is a
QL-observable. A statistical distribution of random oscillations at
the precognitive time scale is represented by its covariance
operator. In this way the brain produces a symmetric positively
defined operator.  By scaling there is obtained the operator $\rho$
which has all properties of the von Neumann density operator, i.e.,
it also has the unit trace. This is the QL density operator. After
this the brain is ready to find the QL-approximation of the
classical statistical average $<f>:$
$$
<f> \approx \rm{Tr} \; \rho A.
$$
We have shown in Khrennikov (2006d) that the classical average, the
statistical sum with respect to the random oscillations at the time
scale $s_{\rm{pc}},$  is approximated by the trace QL-average and
the precision of the QL-approximation is of the magnitude $\kappa$
which is given by (\ref{PC}). Thus if the parameter $\kappa$ is very
small the brain does not lose too much information. This is
practically the CL-computation. But if $\kappa$ is rather large,
then the brain works in a nonclassical regime. One may say (as von
Neumann would like) that in such a regime the brain uses {\it
nonclassical logic.} Huge amounts of information are permanently
neglected. But this does not generate a kind of chaos. Information
is neglected in a consistent way.\footnote{We have a simple picture
of arising nonclassical, in particular, quantum logical structures.
These are systems for processing of incomplete information. The
classical logic would  not be violated if we were able to collect
and process complete sets of information. However, sometimes we are
not able. Therefore we develop special systems for processing of
incomplete sets of information.}

As was pointed out a few times, such a QL-processing of information
save a lot of computational resources. It might be an important
factor of the natural selection of biological organisms.

\section{Multiplicity of  time scales in brain and quantum-like cognitive
representations}

The main lesson from the experimental and theoretical investigations
on the temporal structure of processes in brain is that there are
various time scales. They  correspond to (or least they are coupled
with) various aspects of cognition. Therefore we are not able to
determine once and for ever the cognitive time scale $t_{\rm{c}}$
(``psychological time''). There are few such scales. We shall
discuss some evident possibilities.

Before to go deeper in the temporal structure of mental processes,
we shall analyze in more detail the multi-scale temporal aspects of
quantum mechanics. Such aspects have never been discussed, because,
on the one hand, it was commonly assumed that quantum mechanics is
complete (this is the Copenhagen interpretation), and, on the other
hand, the quantum formalism is used by only one class of observers
-- human beings. The latter generates the unique observational
(quantum) time scale. However, we can consider a possibility that
there exits a class of observers ("super-clocks civilization") which
use a time scale $t_q^\prime$ which is essentially finer than our
time scale $t_q:$
$$
t_q^\prime << t_q
$$
Suppose that the super-clocks civilization has also created the
quantum representation of information. Of course, their time scale
should not be extremely fine comparing with the prequantum  time
scale $s_{\rm{pq}}:$
$$
s_{\rm{pq}} << t_q^\prime
$$
(we assume that both civilizations -- our and super-clocks -- are
interested in processes at the same prequantum time scale). The
super-clocks civilization would discover the same mathematical
formalism of quantum mechanics. But the presence of deviation from
prequantum reality would be more evident with respect to their time
scale (since $s_{\rm{pq}}$ is the same, but $t_q^\prime$ is smaller
than $t_q,$ the coefficient $\kappa^\prime$ for the super-clocks
civilization is larger than the coefficient $\kappa$ for our
civilization).
 On the one hand, the super-clocks civilization has
a better possibility to find deviations of the incomplete quantum
description from the complete classical description. However, there
might be chosen a strategy to ignore such deviations and still use
the quantum picture of the world. Even if it does not match
precisely with the complete set of information about external world,
it might be, nevertheless, convenient (by computational and
consistency reasons) to proceed with the quantum pictures of
reality.

\medskip

Similar functioning with a few time scales of observation (in fact,
self-observation) can be present in the brain. How can we find those
scales?

It is well known, see, e.g., Brazier (1970), that there are well
established time scales corresponding to the alpha, beta, gamma,
delta, and theta waves. Let us consider these time scales as
different cognitive scales. There is one technical deviation from
the QL-scheme which was discussed above. We cannot determine
precisely definite cognitive times corresponding to these scales.
The scales are defined by ranges of frequencies and hence ranges of
scaling times.

For the alpha waves we choose its upper limit frequency, 12 Hz, and hence the
$t_{\rm{c}, \alpha} \approx 0.083$ sec. For the beta waves we consider (by taking upper bounds of
frequency ranges) three different time scales:
15 Hz, $t_{\rm{c}, \beta, \rm{low}} \approx 0.067$ sec. -- low beta waves,
18Hz, $t_{\rm{c}, \beta} \approx 0.056$ sec. -- beta waves,
23 Hz $t_{\rm{c}, \beta, \rm{high}} \approx 0.043$ sec. -- high beta waves. For gamma waves
we take the characteristic frequency 40 Hz and hence the time scale
$t_{\rm{c}, \gamma} \approx 0.025$ sec.

The gamma scale is the finest and hence processes represented at
this scale has the highest degree of QL-ness. On the other hand, we
know that gamma waves patterns in the brain are associated with
perception and consciousness. The beta scale is coarser than the
gamma scale and it has less degree of QL-ness in processing of
information.  We know that beta states are associated with normal
waking of consciousness.

The theta waves are even less QL than the alpha waves. They are
commonly found to originate from occipital lobe during periods of
relaxation, with eyes closed but still awake.  They  are involved
into a representation of information with  a high degree of
classicality. And these rhythms are observed during some sleep
states, and in states of quiet focus, for example, meditation,
Aftanas and Golosheykin (2005). However, there are also experimental
evidences that the theta rhythms are very strong in rodent
hippocampi and entorhinal cortex during learning and memory
retrieval. We can just speculate that learning needs using of an
essentially more detailed information representation. Thus learning
(or at least a part of it) is less QL and hence more CL. The same we
can say about memory retrieval. It also needs more complete,
CL-representation of information. Large body of evidence, Buzsaki
(2005), indicates that theta-rhythms are used in spatial learning
and navigation. Here we present the same reasons: such tasks are
based on CL-representation of information.

Finally, we consider delta waves. Comparing with the highest scale
-- the gamma scale, the delta time scale is extremely rough. This
induces a low degree of QL-ness. This is the state of deep
sleep.\footnote{The phenomena of sleep and dreaming are extremely
complicated. We do not plan to study them  in this paper.}

Although we still did not come to the difficult problem, namely,
determination of the precognitive time scale, we can, nevertheless,
compare the degree of QL-ness of various time scales.

Our choice of the precognitive time scale will be motivated by so
called {\it Taxonomic Quantum Model}, see Geissler et al (1978),
Geissler and Puffe (1982), Geissler (1983, 85, 87,92), Geissler and
Kompass (1999, 2001), Geissler, Schebera, and Kompass (1999),  for
representation of cognitive processes in the brain (which was
developed on the basis of the huge experimental research on
time-mind relation, see also Klix and van der Meer (1978),
Kristofferson (1972, 80, 90),  Bredenkamp (1993), Teghtsoonian
(1971). In the following section we recall briefly the main features
of this model.

\section{Taxonomic quantum model}

There could be presented a portion of good criticism against
starting from EEG bands. Indeed, this band structure is one of the
few indications that directly point to behaviorally relevant
physiological properties. Physiologists suggesting the definitions
had a good intuition. However, that these definitions depend on
behavioral information is shown by enormous individual differences
in the band structures that can be defined only on a behavioral
basis. To some degree this concerns also the general band structure.
Because of individual differences, alpha is often restricted to the
"common" range which is too short to be theoretically fully
relevant. Definitions often go only from 9 to 12 Hz.  Most careful
investigators (earliest Livanov) defined the band by the range 7.5
to 13.5 Hz.

Therefore we propose to start with Taxonomic Quantum Model (TQM),
Geissler et al (1978), Geissler and Puffe (1982), Geissler (1983,
85, 87,92), Geissler and Kompass (1999, 2001), Geissler, Schebera,
and Kompass (1999). Why do we propose to use
 TQM for start of theory instead of, say,
some characteristic physiological parameters such as neuronal
refractoriness, transmission times, coupling strength etc.? In my
view, the reason is that {\it the only basis for interpreting
physiological facts of brain processes are psychophysical
(behavioral) observations, either based on motor reactions of
conscious beings or verbal reports on conscious events.} This was
the main way of thinking of von Bekesy (1936). Of course, many of
the functional statements of physiologists have the same basis. For
our purpose, this statement is absolutely essential, because a
coherent account of temporal properties of brain activity must not
only be related to behavioral observations, but it must be based on
temporal invariants extracted by a coherent theoretical account of
behavioral observations, and only these can provide the guideline to
find the proper physiological correspondences.

The best short cut to the approach is through the history of its
emergence: The first impulse towards a taxonomic turn arose in the
early 1970s from the discontent of  Geissler, see, e.g., Geissler et
al (1978), with the fact that in simple psychophysical tasks data
could indistinguishably be fitted to models resorting to widely
differing, often enough even contradicting, assumptions. In his
research in visual recognition, to circumvent this difficulty,
Geissler introduced a technique of "chronometric cross-task
comparison. " The main idea was to disambiguate models by temporal
parametrization, thereby postulating invariance of time parameters
under variation of stimulus parameters and task constraints (see
e.g. Geissler et al. (1978) and Geissler and Puffe (1982)). At that
time another research group at the same institute did something
similar by fitting latencies in standardized reasoning tasks to
predicted numbers of operations, e.g., Klix and van der Meer (1978).
The estimates from the two lines of studies yielded a surprising
picture: There seemed to exist small "bands" of operation times
centering at around 55, 110 and 220 ms, thus exhibiting
near-doubling relations. As a datum from the literature which fitted
into this regularity the asymptotic value of 36.5 ms determined by
Kristofferson (1972), see also  Kristofferson (1980, 90),  came to
mind which up to the first decimal is 1/3 of 110 ms. Taken together,
these four values suggested a system of "magic numbers". Herein a
period of 110 ms represents something like a "prototype duration"
from which the rest of periods derives by either integer division or
multiplication. From various fit procedures for step lengths,
Buffart and Geissler came up with an largest common denominator
(l.c.d.) of 9.13 ms (see Geissler, 1985) showing a standard
deviation of 0.86 ms across individuals. It turned out that the four
above-mentioned periods, although partly many times larger than this
small period, can be represented as integer multiples of it, with
nearly absolute precision: $4 \times 9.13 = 36.5; 6  \times 9.13 =
54.8; 12  \times 9.13 = 109.6; 24  \times 9.13 = 219.1.$ Of course,
this might have been some strange coincidence. Yet, later,
chronometric analysis seemed to support a modular unit of some 9 ms
(see Geissler (1985); Puffe (1990); Bredenkamp (1993). Further
investigations justify a modified assumption about quantal graining:

\medskip

{\it Regression yields the largest common denominator (l.c.d.) 4.6
ms, which is nearly exactly one half of 9.13 ms.}

\medskip

Note that, in terms of hypothetical quanta, a period of such duration represents the
next smaller candidate of a "true" elementary "time quantum" which is compatible
 with the recognition data. In the following, let us adopt provisionally the
("ideal") value of
$$
Q_0 = 4.565 \rm{ms}
$$  for this time quantum hypothesis.

The solution TQM offers to these seeming contradictions, see
Geissler (1987, 92, 85) can be considered as a generalization or at
least an analogue of the psychophysical principle of relative-range
constancy.   According to Teghtsoonian (1971), this principle
expresses itself in the fact that for all sensory continua, in terms
of output magnitudes, the ratio of the largest to the smallest
quantity is a constant of around 30. About the same value is
obtained from the so-called Subjective Weber Law.

The generalization of the principle in the realm of quantal timing
is the quantal-range constraint. To see how this analogue reads, consider
first the assumed smallest period $Q_0.$ For integer multiples
$n \times Q_0,$
consistency with the relative range constraint implies $n \leq M,$ with $M$
being a constant of the hypothetical value 30. It follows that periods of durations in
excess of $30 \times Q_0 \approx 137$ ms cannot be represented within this smallest
possible range. To account for such periods, we have to assume larger ranges
with correspondingly larger admissible smallest quantal periods to be operative.
To retain consistency with the time quantum assumption, these periods must be
integer multiples of $Q_0$ or, formally,
\begin{equation}
\label{TR}
Q_q = q \times Q_0
\end{equation}
with integer $q$ must hold. Thus, in general, the maximum extension
of any quantal of periods $T_i$ belonging to it is given by
$q \times Q_0 \leq T_i \leq M \times q \times Q_0.$
Note that the lower bound $q \times Q_0$
also defines the smallest possible distance between admissible periods within a range.
For this reason we will speak of it as the quantal resolution within a given range.
Of course, in the actual development, this abstract definition resulted from a
variety of empirical relationships suggesting a range ordering of quantal periods
with upper bounds maximally at 30 times the value of quantal resolution.

TQM does not exclude the possibility that there can be found smaller
characteristic time scales, e.g., $Q_0/30.$

\section{Precognitive time scale}

We choose $Q_0$ as the unit of the precognitive time:
\begin{equation} \label{PC} s_{\rm{pc}}=Q_0=4.6 ms
\end{equation}
This corresponds to frequencies $\approx 220$ Hz. Under such an
assumption about the precognitive scale we can find the measure of
QL-ness for different EEG bands. For the alpha scale, we have
$$
\kappa_\alpha= \frac{Q_0}{t_{\rm{c}, \alpha}}\approx 0.055.
$$
For the beta scales, we have:
$$
\kappa_{\rm{c}, \beta, \rm{low}}= \frac{Q_0}{t_{\rm{c}, \beta, \rm{low}}}\approx 0.069;\;
\kappa_{\rm{c}, \beta}=\frac{Q_0}{t_{\rm{c}, \beta}}\approx 0.082;\;
\kappa_{\rm{c},\beta,\rm{high}}= \frac{Q_0}{t_{\rm{c},\beta,\rm{high}}}\approx 0.107.
$$
For the gamma scale we have:
$$
\kappa_\gamma= \frac{Q_0}{t_{\rm{c}, \gamma}}\approx 1.84.
$$
Thus QL-ness of processing of information increases. ``Thinking
through the alpha waves'' is more likely processing of information
by ordinary computer. Not so much information is neglected.
Therefore the information processing is not so tricky: there is no
need to manipulate with extremely incomplete information in the
consistent way. ``Thinking through the gamma waves'' is similar to
processing of information by an analogue of quantum computer --
QL-computer, see Khrennikov (2006a). Such an information processing
is very tricky: permanent informational cuts, but in the consistent
QL-way. Finally, we come to the theta and delta scales. For the
theta scale $t_{\rm{c}, \theta}= 0.125$ sec. Thus
$$
\kappa_\theta= \frac{Q_0}{t_{\rm{c},\theta}}\approx 0.037.
$$
And for the delta scale $t_{\rm{c}, \delta}= 0.5$ sec and hence:
$$
\kappa_\delta= \frac{Q_0}{t_{\rm{c},\delta}} \approx 0.009.
$$
Here the difference between the biological QL-processing of information  in the brain
and the CL-processing (as in models of artificial intelligence)  is practically
negligible.

We now compare our QL-scales of time with the "quantum scales" which
were chosen in Khrennikov (2006d):
\begin{equation}
\label{PCX} s_{\rm{pc}}\approx 10^{-3} \; \rm{sec}, \;  \;  \;
s_{\rm{pq}}\approx 10^{-44} \; \rm{sec}.
\end{equation}
\begin{equation}
\label{PCX1} t_{\rm{c}}= 30 Q_0\approx 10^{-1} \; \rm{sec}, \;  \;
\; t_{\rm{q}}\approx 10^{-21} \; \rm{sec}.
\end{equation}
Thus our model is based on macroscopic time scales, in the
opposition to really quantum reductionist models.

If we follow TQM in more detail then we should consider a
possibility that in the brain there exist a {\it hierarchy of
precognitive times,} i.e., the above model with one fixed
precognitive time given by (\ref{PC}) was oversimplified. From the
point of view of TQM each $Q_q$ given by (\ref{TR}) could serve as
the basis of a precognitive time scale. We obtain a picture of
extremely complex QL-processing of information in the brain which is
based of the huge multiplicity of various precognitive/cognitive
scales.

In this framework the notion
``precognitive'' loses its absolute meaning. The notions
``precognitive''/``cognitive'' become
relative with respect to a concrete psychological function
(cognitive task). Moreover, a time scale which is precognitive
for one psychological function can be at the same time cognitive for another.

But the crucial point is that the same cognitive time scale, say $t_{\rm{c}},$ can
have a number of different precognitive scales:
$$
Q_{q_1} \leq ... \leq Q_{q_m}.
$$
Each pair of scales
$$
(Q_{q_1}, t_{\rm{c}}),..., (Q_{q_m}, t_{\rm{c}})
$$
induces its own QL-representation of information. Therefore the same
$t_{\rm{c}}$-rhythm can be involved in the performance of a few
different psychological functions.

The final message from TQM is that the cognitive time $t_{\rm{c}}$
scale should be based on an integer multiplier of the time quant
$Q_0:$
\begin{equation}
\label{MP}
t_{\rm{c}}= N Q_0.
\end{equation}
In such a model we can totally escape coupling with directly defined
different EEG bands, alpha, beta, gamma,... We shall use only
behaviorally defined time scales. The Weber law gives us the
restriction to the value of the multiplier: $N \leq 30.$

\bigskip

{\bf References}

Aerts, D. and  Aerts, S., editors, 2007. Optimal Observation: From
Quantum Mechanics to Signal Analysis (Einstein Meets Magritte: An
Interdisciplinary Reflection on Science, Nature, Art, Human Action
and Society). Springer, Berlin-Heidelberg.

Aftanas, L., Golosheykin, S., (2005). Impact of regular meditation
practice on EEG activity at rest and during evoked negative
emotions. Int J Neurosci.115(6), 893-909.

Albert, D. Z., Loewer, B., 1988. Interpreting the many worlds
interpretation. Synthese 77, 195-213.

Albert, D. Z., 1992. Quantum mechanics and experience. Cambridge,
Mass.: Harvard Univ. Press.

Amit, D., 1989. Modeling Brain Function. Cambridge Univ. Press,
Cambridge.

Ashby, R., 1952,  Design of a brain. Chapman-Hall, London.

Barrett,  J. A., 1999.  The quantum mechanics of minds and worlds.
Oxford Univ. Press, Oxford.

Bechtel, W., Abrahamsen,  A.,  1991. Connectionism and the mind.
Basil Blackwell, Cambridge.

Bohm, D., Hiley,   B., 1993.  The undivided universe: an ontological
interpretation of quantum mechanics. Routledge and Kegan Paul,
London.

Brazier, M. A. B., 1970. The Electrical Activity of the Nervous
System. London: Pitman

Bredenkamp, J. (1993). Die Verknupfung verschiedener
Invarianzhypothesen im Bereich der Gedachtnispsychologie.
Zeitschrift fur Experimentelle und Angewandte Psychologie, 40,
368-385.

Buzsaki, G., 2005. Theta rhythm of navigation: link between path
integration and landmark navigation, episodic and semantic memory.
Hippocampus. 15(7), 827-40.

Choustova, O., 2004. Bohmian mechanics for financial processes. J.
Modern Optics 51, n. 6/7, 1111.

Conte, E., Todarello, O., Federici,  A.,  Vitiello, T., Lopane,  M.,
Khrennikov,  A. Yu., Zbilut J. P., 2007. Some remarks on an
experiment suggesting quantum-like behavior of cognitive entities
and formulation of an abstract quantum mechanical formalism to
describe cognitive entity and its dynamics. CHAOS SOLITONS and
FRACTALS 31 (5), 1076-1088.
http://xxx.lanl.gov/abs/quant-ph/0307201.

Cossart, R., Aronov, D.  and  Yuste, R. (2003) Nature 423, 283-288.

Deutsch, D., 1997. The Fabric of Reality. How much can our four
deepest theories of the world explain? Publisher Allen Lane, The
Penguin Press.

Donald, M. J., 1996. On many-minds interpretation of quantum
mechanics. Preprint.

Donald, M. J., 1990. Quantum theory and brain. Proc. Royal Soc. A
427, 43-93.

Donald, M. J., 1995. A mathematical characterization of the physical
structure of observers. Found. Physics 25/4, 529-571.

Eliasmith, C., 1996.  The third contender: a critical examination of
the dynamicist theory of cognition.  Phil. Psychology 9(4), 441-463.

Geissler, H.-G., Klix, F., and Scheidereiter, U. (1978). Visual
recognition of serial structure: Evidence of a two-stage scanning
model. In E. L. J. Leeuwenberg and H. F. J. M. Buffart (Eds.),
Formal theories of perception (pp. 299-314). Chichester: John Wiley.

Geissler, H.-G. and Puffe, M. (1982). Item recognition and no end:
Representation format and processing strategies. In H.-G. Geissler
and Petzold (Eds.), Psychophysical judgment and the process of
perception (pp. 270-281). Amsterdam: North-Holland.

Geissler, H.-G. (1983). The Inferential Basis of Classification:
From perceptual to memory code systems. Part 1: Theory. In H.-G.
Geissler, H. F. Buffart, E. L. Leeuwenberg, and V. Sarris (Eds.),
Modern issues in perception (pp. 87-105). Amsterdam: North-Holland.

Geissler, H.-G. (1985). Zeitquantenhypothese zur Struktur
ultraschneller Gedachtnisprozesse. Zeitschrift fur Psychologie, 193,
347-362. Geissler, H.-G. (1985b). Sources of seeming redundancy in
temporally quantized information processing. In G. d'Ydewalle (Ed.),
Proceedings of the XXIII International Congress of Psychology of the
I.U.Psy.S., Volume 3 (pp. 199-228). Amsterdam: North-Holland.

Geissler, H.-G. (1987). The temporal architecture of central
information processing: Evidence for a tentative time-quantum model.
Psychological Research, 49, 99-106. Geissler, H.-G. (1990).
Foundations of quantized processing. In H.-G. Geissler (Ed.),
Psychophysical explorations of mental structures (pp. 193-210).
Gottingen, Germany: Hogrefe and Huber Publishers.

Geissler, H.-G. (1992). New magic numbers in mental activity: On a
taxonomic system for critical time periods. In H.-G. Geissler, S. W.
Link, and J. T. Townsend (Eds.): Cognition, information processing
and psychophysics (pp. 293-321). Hillsdale, NJ: Erlbaum.

Geissler, H.-G. and Kompass, R. (1999). Psychophysical time units
and the band structure of brain oscillations. 15th Annual Meeting of
the International Society for Psychophysics, 7-12.

Geissler, H.-G., Schebera, F.-U., and Kompass, R. (1999).
Ultra-precise quantal timing: Evidence from simultaneity thresholds
in long-range apparent movement. Perception and Psychophysics, 6,
707-726.

Geissler, H.-G., and Kompass, R. (2001). Temporal constraints in
binding? Evidence from quantal state transitions in perception.
Visual Cognition, 8, 679-696.

Hameroff,  S., 1994. Quantum coherence in microtubules. A neural
basis for emergent consciousness?  J. of Consciousness Studies, 1,
91-118.

Hameroff,   S., 1998. Quantum computing in brain microtubules? The
Penrose-Hameroff Orch Or model of consciousness. Phil. Tr. Royal
Sc., London A, 1-28.

Healey, R., 1984. How many worlds? Nous 18, 591-616.

Hiley,  B.,  Pylkk\"anen, P., 1997.  Active information and
cognitive science -- A reply to Kiesepp\"a. In: Brain, mind and
physics.  Editors: Pylkk\"anen, P., Pylkk\"o, P., Hautam\"aki, A.
IOS Press, Amsterdam.

Hiley, B., 2000. Non-commutavive geometry, the Bohm interpretation
and the mind-matter relationship. Proc. CASYS 2000, Liege, Belgium.

Hopfield,  J. J., 1982. Neural networks and physical systems with
emergent collective computational abilities. Proc. Natl. Acad. Sci.
USA 79,  1554-2558.

Hoppensteadt, F. C., 1997.  An introduction to the mathematics of
neurons: modeling in the frequency domain.  Cambridge Univ. Press,
New York.

Ikegaya, Y., Aaron, G., Cossart, R., Aronov, D., Lampl, I., Ferster,
D. and Yuste, R. (2004) Science 304, 559-564.

Jibu, M., Yasue,  K., 1992. A physical picture of Umezawa's quantum
brain dynamics. In  Cybernetics and Systems Research, ed. R. Trappl,
World Sc., London.

Jibu, M.,  Yasue,  K., 1994.  Quantum brain dynamics and
consciousness. J. Benjamins Publ. Company, Amsterdam/Philadelphia.

Kerr, J. N., Greenberg, D. and Helmchen, F. (2005) Proc. Natl. Acad.
Sci. U. S. A 102, 14063-14068.

Khrennikov, A. Yu., 1999. Classical and quantum mechanics on
information spaces with applications to cognitive, psychological,
social and anomalous phenomena. Found. Phys. 29,  1065-1098.

Khrennikov,  A. Yu., 2000.  Classical and quantum mechanics on
$p$-adic trees of ideas. BioSystems   56, 95-120.

Khrennikov, A. Yu., 2002. On cognitive experiments to test
quantum-like behaviour of mind. Rep. V\"axj\"o Univ.: Math. Nat. Sc.
Tech., N 7; http://xxx.lanl.gov/abs/quant-ph/0205092.

Khrennikov,  A. Yu., 2003. Quantum-like formalism for cognitive
measurements. Biosystems 70, 211-233.

Khrennikov, A. Yu., 2004. Information Dynamics in Cognitive,
Psychological, Social, and Anomalous Phenomena (Fundamental Theories
of Physics). Springer, Berlin-Heidelberg.

Khrennikov, A. Yu., 2006a. Quantum-like brain: "Interference of
minds" Biosystems 84 (3),  225-241.

Khrennikov, A. Yu., 2005a. A pre-quantum classical statistical model
with infinite-dimensional phase space. J. of Physics A, Math. and
General 38 (41), 9051-9073.

Khrennikov, A. Yu., 2005b. Generalizations of quantum mechanics
induced by classical statistical field theory. Foundations of
Physics Letters  18 (7),  637-650.

Khrennikov, A. Yu., 2006b. On the problem of hidden variables for
quantum field theory. Nuovo Cimento B  121 (5), 505-521.

Khrennikov, A. Yu., 2006c. Nonlinear Schrodinger equations from
prequantum classical statistical field theory. PHYSICS LETTERS A 357
(3), 171-176,

 Khrennikov, A. Yu., 2006d. To quantum mechanics through random
fluctuations at the Planck time scale.
http://www.arxiv.org/abs/hep-th/0604011

Klix, F., and van der Meer, E. (1978). Analogical reasoning - an
approach to mechanisms underlying human intelligence performances.
In F. Klix (Ed.), Human and artificial Intelligence (p. 212).
Berlin: Deutscher Verlag der Wissenschaften.

Kristofferson, M. W. (1972). Effects of practice on
character-classification performance. Canadian Journal of
Psychology, 26, 540-560.

Kristofferson, A. B. (1980). A quantal step function in duration
discrimination. Perception and Psychophysics, 27, 300-306.

Kristofferson, A. B. (1990). Timing mechanisms and the threshold for
duration. In Geissler, H.-G. (Ed., in collaboration with M. H.
Muller  and  W. Prinz), Psychophysical explorations of mental
structures (pp. 269-277). Toronto: Hogrefe  and  Huber Publishers.

Lockwood, M., 1989. Mind, Brain and Quantum. Oxford, Blackwell.

Lockwood, M., 1996. Many minds interpretations of quantum mechanics.
British J. for the Philosophy of Sc. 47/2, 159-88.

Loewer, B., 1996. Comment on Lockwood. British J. for the Philosophy
of Sc. 47/2, 229-232.

Luczak, A., Bartho, P., Marguet, S. L., Buzsaki, G., Hariis, K.D,
2007. Neocortical spontaneous activity in vivo: cellular
heterogeneity and sequential structure. Preprint of CMBN, Rutgers
University.

MacLean, J. N., Watson, B. O., Aaron, G. B. and Yuste, R. (2005)
Neuron 48, 811-823.

Mao, B. Q., Hamzei-Sichani, F., Aronov, D., Froemke, R. C. and
Yuste, R. (2001) Neuron 32, 883-898.

Orlov, Y. F., 1982.  The wave logic of consciousness: A hypothesis.
Int. J. Theor. Phys. 21, N 1, 37-53.

Penrose, R., 1989. The emperor's new mind. Oxford Univ. Press,
New-York.

Penrose, R., 1994.  Shadows of the mind. Oxford Univ. Press, Oxford.

Penrose, R., 2005. The Road to Reality : A Complete Guide to the
Laws of the Universe. Knopf Publ.

Plotnitsky, A., 2001. Reading Bohr: Complementarity, Epistemology,
Entanglement, and Decoherence. Proc. NATO Workshop Decoherence and
its Implications for Quantum Computations, Eds. A.Gonis and
P.Turchi, p.3--37, IOS Press, Amsterdam.

Plotnitsky, A., 2002. Quantum atomicity and quantum information:
Bohr, Heisenberg, and quantum mechanics as an information theory,
Proc. Conf.  Quantum theory: reconsideration of foundations, ed: A.
Yu. Khrennikov, Ser. Math. Modelling
 2, 309-343, V\"axj\"o Univ. Press,  V\"axj\"o.

Plotnitsky, A., 2007. Reading Bohr: Physics and Philosophy
(Fundamental Theories of Physics). Springer, Berlin-Heidelberg.

Pylkk\"anen, P., 2006. Mind, Matter and the Implicate Order (The
Frontiers Collection). Springer, Berlin-Heidelberg.

Puffe, M. (1990). Quantized speed-capacity relations in short-term
memory. In H.-G. Geissler (Ed., in collaboration with M. H. Muller
and W. Prinz), Psychophysical exploration of mental structures (pp.
290-302). Toronto: Hogrefe and Huber Publishers.

Sanchez-Vives, M. V. and McCormick, D. A. (2000) Nat. Neurosci. 3,
1027-1034.

 Shu, Y. S., Hasenstaub, A. and McCormick, D. A. (2003) Nature 423, 288-293.

Stapp, H. P.,  1993. Mind, matter and quantum mechanics.
Springer-Verlag, Berlin-New York-Heidelberg.

Strogatz,  S. H., 1994.  Nonlinear dynamics and chaos with
applications to physics, biology, chemistry, and engineering.
Addison Wesley, Reading, Mass.

Svozil, K., 2006. Quantum Logic (Discrete Mathematics and
Theoretical Computer Science). Springer, Berlin-Heidelberg.

Teghtsoonian, R. (1971): On the exponents in Stevens' law and on the
constant in Ekman's law. Psychological Review, 78, 71 - 80.

van Gelder, T., Port, R., 1995. It's about time: Overview of the
dynamical approach to cognition. in  Mind as motion: Explorations in
the dynamics of cognition. Ed.: T. van Gelder, R. Port. MITP,
Cambridge, Mass, 1-43.

van Gelder, T., 1995. What might cognition be, if not computation?
J. of Philosophy 91, 345-381.

Vitiello, G., 2001.  My double unveiled - the dissipative quantum
model of brain. J. Benjamins Publ. Company, Amsterdam/Philadelphia.

von Bekesy, G. (1936). Uber die Horschwelle und Fuhlgrenze langsamer
sinusformiger Luftdruckschwankungen. Annalen der Physik, 26,
554-556.

von Neumann, J.,  1955. Mathematical foundations of quantum
mechanics. Princeton Univ. Press, Princeton, N.J..

Whitehead, A. N., 1929. Process and Reality: An Essay in Cosmology.
Macmillan Publishing Company, New York.

Whitehead, A. N., 1933.  Adventures of Ideas.  Cambridge Univ.
Press, London.

Whitehead, A. N., 1939. Science in the modern world. Penguin,
London.

\end{document}